\UseRawInputEncoding

\documentclass{vgtc}                          

\graphicspath{{figures/}{pictures/}{images/}{./}} 

\usepackage{times}                     

\usepackage{tabu}                      
\usepackage{booktabs}                  
\usepackage{lipsum}                    
\usepackage{mwe}                       

\usepackage{mathptmx}                  

\onlineid{0}

\vgtccategory{Research}

\vgtcinsertpkg

\title{Fields, Bridges, and Foundations: How Researchers Browse \\ Citation Network Visualizations}

\author{Kiroong Choe\\ %
        \scriptsize Seoul National University %
\and Eunhye Kim\\ %
     \scriptsize Seoul National University %
\and Sangwon Park\\ %
    \scriptsize Seoul National University %
\and Jinwook Seo\\ %
    \scriptsize Seoul National University }%

\teaser{
  \centering
  \includegraphics[width=\linewidth]{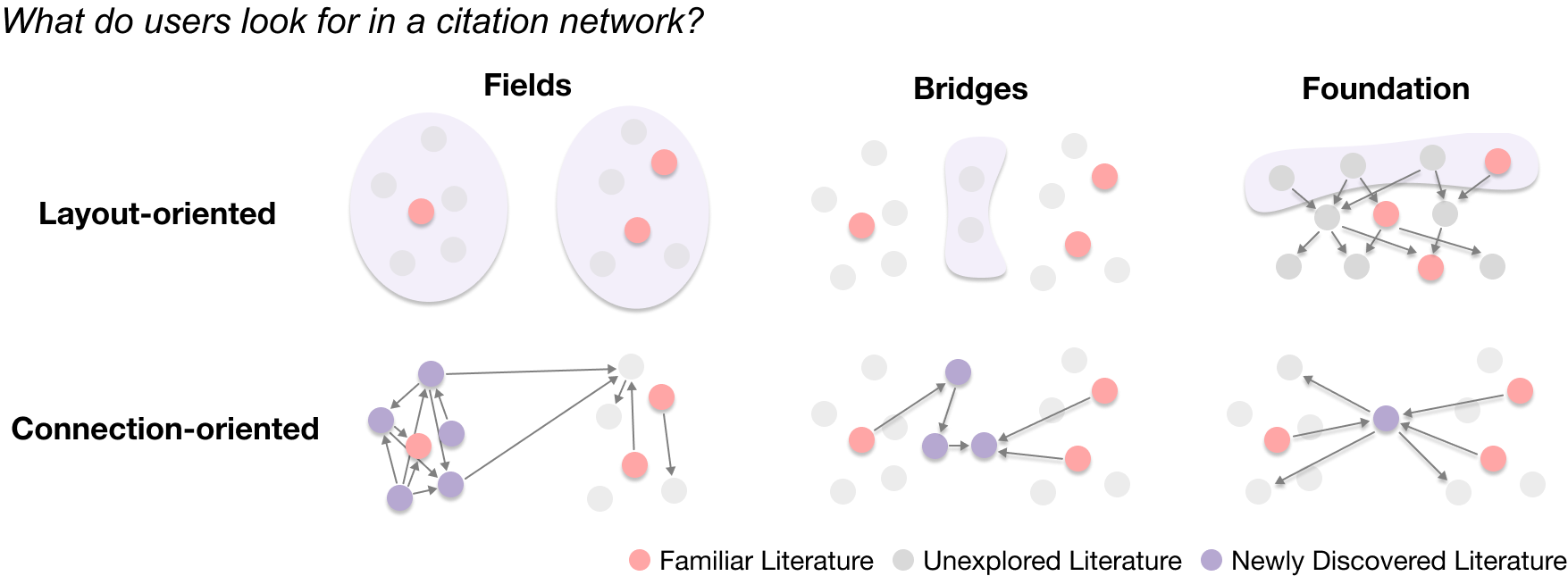}
  \caption{We identified six patterns that researchers utilize to browse citation networks and discover papers of interest. Component-wise, these patterns can be classified to: Field (\textit{i.e.,} related papers on a single research topic), Bridge (\textit{i.e.,} logical connections between papers or topics), and Foundation (\textit{i.e.,} stages in the broad development of research). For each component, there were two different perspectives: layout-oriented or connection-oriented. Our analysis suggests that researchers generally preferred the layout-oriented perspective for its intuitiveness, but papers identified through the connection-oriented perspective were typically more useful.}
  \label{fig:teaser}
}

\abstract{
Visualizing citation relations with network structures is widely used, but the visual complexity can make it challenging for individual researchers trying to navigate them. We collected data from 18 researchers with an interface that we designed using network simplification methods and analyzed how users browsed and identified important papers. Our analysis reveals six major patterns used for identifying papers of interest, which can be categorized into three key components: Fields, Bridges, and Foundations, each viewed from two distinct perspectives: layout-oriented and connection-oriented. The connection-oriented approach was found to be more reliable for selecting relevant papers, but the layout-oriented method was adopted more often, even though it led to unexpected results and user frustration. Our findings emphasize the importance of integrating these components and the necessity to balance visual layouts with meaningful connections to enhance the effectiveness of citation networks in academic browsing systems.
} 


\begin{document}


\firstsection{Introduction}

\maketitle


Understanding the relationships between academic literature is vital across all research disciplines, with particular complexity in multidisciplinary fields such as Human-Computer Interaction (HCI), which bridges computer engineering, design, and social sciences.~\cite{oulasvirta2016hci, liu2014chi}. Researchers engage in browsing individual papers or topic-oriented exploration to achieve these goals~\cite{choe2021papers101, sultanum2020understanding, kang2022threddy, kang2023synergi}. They typically begin with a few relevant papers for a specific research topic, and these papers are generally called \textit{``seed papers''}. Seed papers help researchers progressively expand their knowledge in related areas, identify additional significant papers, and develop a comprehensive perspective on related work.


Citation relationships are a fundamental component of academic databases, providing essential insights into the landscape and trends of a research field. Visualizing citations with networks, representing papers as nodes and citation relationships as links, is an intuitive and widely used method for illustrating these connections. However, it can be challenging to browse a citation network with ease due to the high visual complexity. As a result, most studies utilize citation networks for specific tasks such as detecting macro trends or anomalies using analysis methods~\cite{nakazawa2015visualization, shen2006biblioviz, Alfradi, emrouznejad2017state, felizardo2012visual}. However, tailoring citation networks for a specific goal cannot support general browsing needs, which may have diverse goals or may start without a specific goal or task. While systems designed to browse literature related to specific topics (e.g., Connected Papers \cite{behera2023visual}) do show the citations among recommended relevant papers, the system selectively presents references based on specific criteria, and this limits users from being able to explore further aspects of the network.

One way to enable manual browsing of citation network visualizations is to simplify the graph before user interaction and to provide details on demand. Techniques to summarize graphs have alleviated the cognitive load of interpreting graphs by simplifying their substructures. For example, Dunne~\cite{dunne2013motif} proposed representing three motifs---fans, connectors, and (near) cliques---in a glyph form to effectively visually summarize complex graphs. Similarly, Koutra~\cite{koutra2015summarizing} enabled the detection and display of (near) cliques, bipartite cores, stars, and chains within the graph. While these simplification methods can assist in exploring citation graphs, the relevance of different graph components can vary depending on individual researchers' needs and perspectives. Without understanding what researchers consider important or irrelevant, it remains challenging to determine which components should be emphasized, compressed, or hidden.

To investigate which components researchers are interested in, we designed and deployed an interface that allowed 18 study participants to browse citation graphs based on their selected seed papers, through which we identified six major patterns (\Cref{fig:teaser}). There were two distinct perspectives: layout-oriented and connection-oriented. The connection-oriented perspective yielded a more reliable selection of relevant papers than the layout-oriented perspective. However, many participants preferred the layout-oriented perspective, assuming the network layout followed an intuitive and straightforward interpretive rule. This assumption often resulted in unexpected outcomes and frustration when it did not align with the actual network structure.
We highlight the need to bridge the gap between the visual layout of citation networks and their meaningful connection structures.

\section{Method}


\subsection{Pilot Study}

Our initial approach to understanding visual structures was to simplify the large citation network by breaking it down around seed papers into communities small enough to be displayed. With the seed papers selected by the pilot study participants, we extracted neighboring papers using the Breadth-First Search algorithm using the Semantic Scholar API~\cite{s2api}. The graphs were then segmented into sufficiently small communities by recursively applying the Louvain community detection algorithm~\cite{de2011generalized}.

We ensured that communities consisted of fewer than 30 nodes to keep them visually easy to perceive, resulting in communities typically being formed with a single seed paper and a tree-like structure of cited papers around it. We tested this approach on three participants as a pilot study. We found that our community structure lacked informational value, and participants were curious about the qualitative differences between nodes included or excluded by the community algorithm. Instead, participants were interested in looking at the connections between multiple seed papers that they had chosen. The pilot study indicated a need to provide larger graphs that display multiple seed papers and their interrelations despite the potential risk of visual complexity.

\subsection{Citation Network Browsing Interface}

\begin{figure*}[t]
 \centering
 \includegraphics[width=\linewidth]{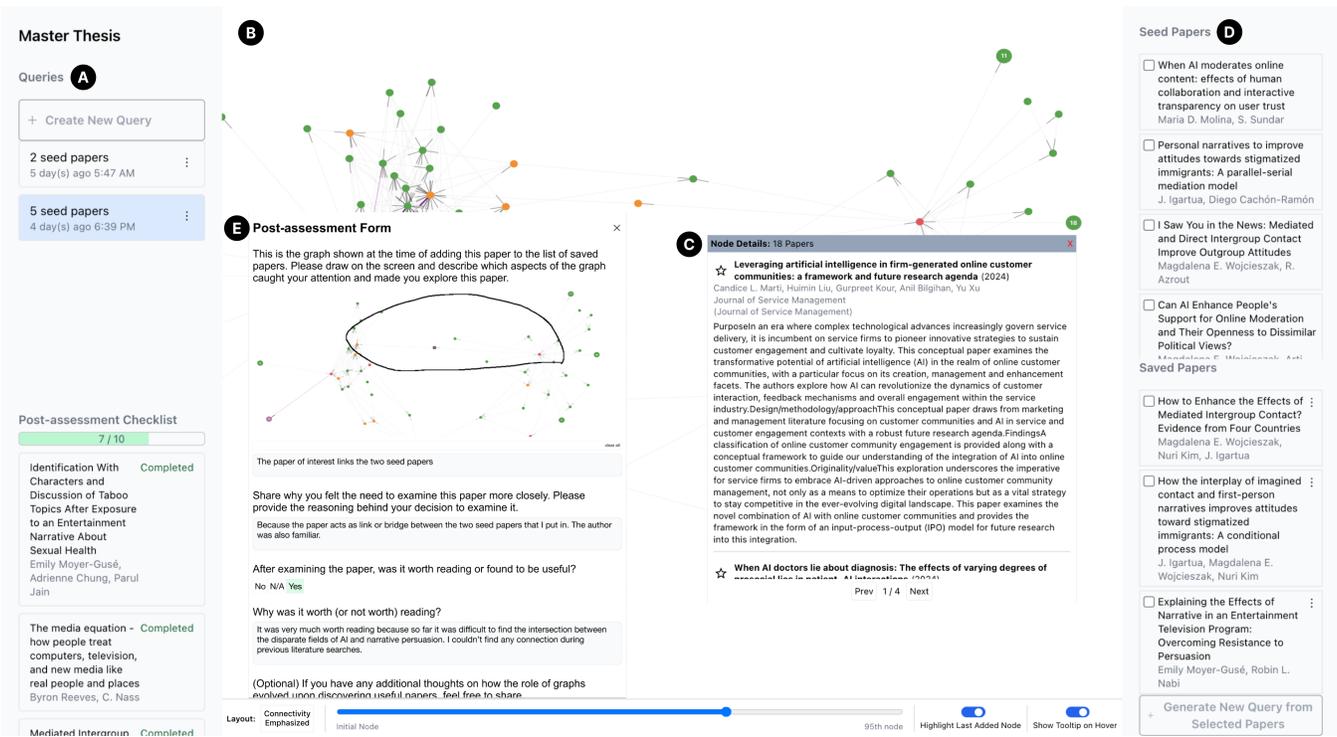}
 \caption{A screenshot of the interface used in the experiment. After inputting seed papers (A), participants were asked to browse the citation network (B, C), identify noteworthy papers (D), and provide feedback on why they were interested in specific parts of the graph and how the papers they found were actually useful (E).}
 \label{fig:interface}
\end{figure*}

\begin{figure}[t]
 \centering
 \includegraphics[width=0.8\linewidth]{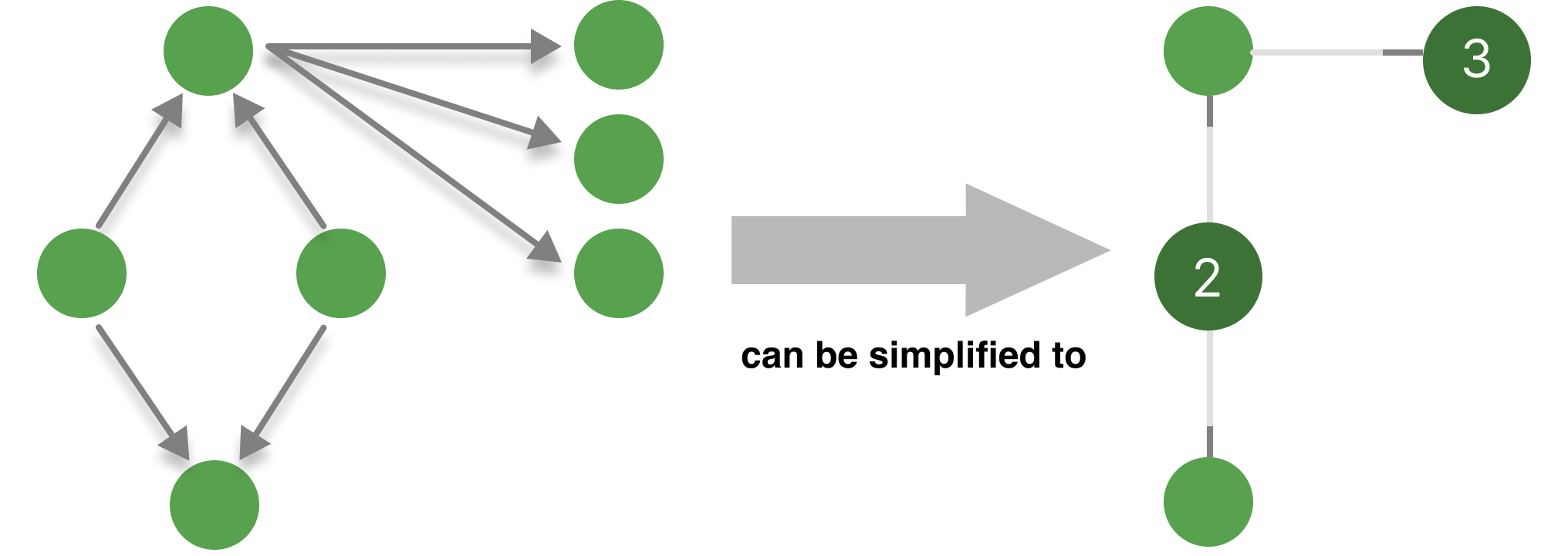}
 \caption{We simplified the graph by using colors on the tip of the link to indicate direction and merging nodes with identical connections.}
 \label{fig:arrow}
\end{figure}

\begin{figure}[t]
 \centering
 \includegraphics[width=\linewidth]{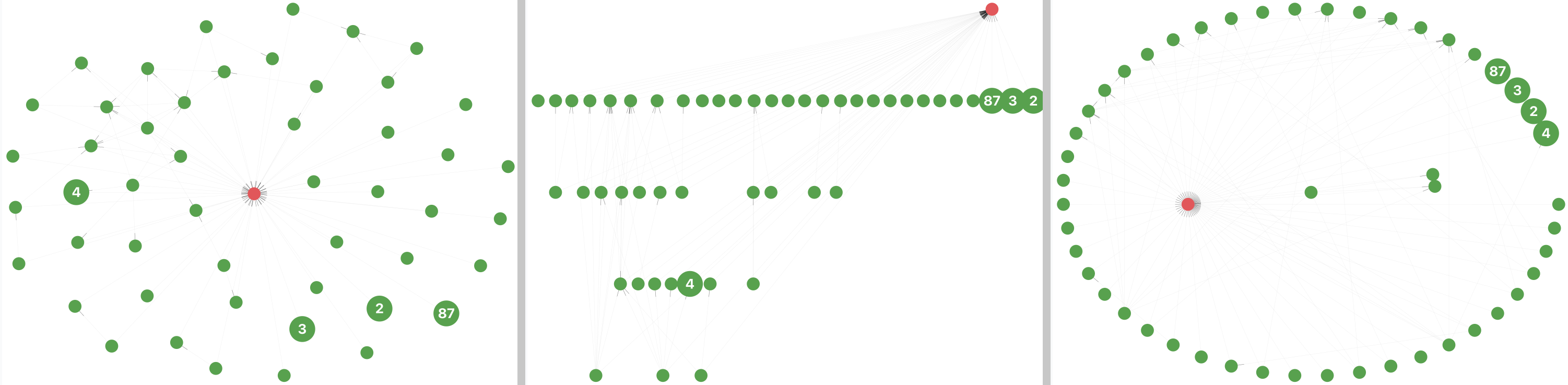}
 \caption{Examples of the three layouts provided to users: force-directed (left), rank-based (center), and circular (right).}
 \label{fig:layouts}
\end{figure}

Based on our pilot study results, we developed an interface that displays a graph consisting of at least 100 densely connected papers and their citation relationships, starting from the seed papers given as input by the user. We designed the interface to fetch nodes in order of their connectivity to the given seed papers. It iteratively expands a set, starting with the seed papers, by adding papers that share the most direct citation links with the papers already in the set. This adjustment addresses participants' desire from the pilot study to see connections between multiple seed papers. Consequently, as new nodes are discovered and added to the seed paper list or previous seed papers are removed by the user, the 100 nodes consisting of the graph also change, allowing users to browse the literature space by creating multiple combinations of seed papers.

We applied graph summarization techniques to reduce visual complexity (\Cref{fig:arrow}). The first technique was to modify arrows in edges that represent citation relationships. Instead of using conventional triangular arrowheads, which can overlap in nodes with a high in-degree, we simplified the arrow by adding color on the tip of the edge to represent citation direction. The second summarization involves merging nodes with the same connection relationships, serving a similar purpose to Dunne’s fan and connector motif simplification~\cite{dunne2013motif}  to avoid displaying excessive leaf nodes. This prevents the layout from spreading too widely across the screen when a paper has multiple references without interconnection. After the summarization, we limited the number of nodes to 100, which was an empirical decision made to avoid excessive visual complexity.

Additionally, the system offers three different layout algorithms (\Cref{fig:layouts}). Layouts play a crucial role in interpreting graphs, and certain inherent relationships within a graph can only be perceived effectively in a certain layout. In addition to the commonly used force-directed and circular layouts in previous research~\cite{li2016evaluation, soni2018perception}, we added the DOT layout, a rank-based arrangement that shows node hierarchy. These layouts were calculated using the GraphViz library’s \textit{``sfdp''}, \textit{``twopi''}, and \textit{``dot''} layout engines~\cite{gansner2009drawing}, and provided to users under the names \textit{``connectivity emphasized,''} \textit{``hierarchy emphasized,''} and \textit{``circular arrangement.''} We intentionally gave users only these names to encourage them to discover each layout's unique roles and opportunities without cognitive effort to understand the layout itself.

\subsection{Task}

After inputting seed papers for their research, participants used the system to explore the citation network. Users could access metadata for each node (\textit{i.e.,} title, authors, citation count, journal, year, abstract) through a popup by clicking on the node of interest. Users could \textit{``star''} noteworthy papers, and they were required to save ten papers for post-assessment. The post-assessment was a procedure of annotating whether the paper saved was related to their research topic and useful enough to be referenced in their research. Users conducted this assessment using their preferred method, whether by reading the abstract, skimming, or thoroughly reading the papers. To ensure sufficient time for the post-assessment, we deployed the system for two days, during which the users could perform the evaluations freely.

The post-assessment consisted of several questions, as shown in \cref{fig:interface}-E. 
At the top, a snapshot of the graph of the moment the paper was saved was given. The first question asked why the user found the particular node interesting in the given graph structure. Users could respond by drawing or marking on the snapshot in addition to providing text answers. 
The second question was a yes or no question about whether the paper was useful to their research. Subsequent questions asked why they thought the paper was (or was not) worth reading and if their thoughts about the network had changed.

Prior to the experiment, participants were required to watch a 20-minute instructional video and read the user manual, which only included a description of how to use the system and the supported features. This was followed by two days of individual time for the post-assessment. After completing the task, a 30-minute interview was conducted for debriefing, where we asked how they used the graph, their goals of using the system and overall satisfaction, and how they interpreted the graph structures. Each participant who completed all procedures received a compensation of 20,000 KRW (about 14.5 USD).

\subsection{Participants and Data Analysis}

We recruited researchers from the HCI and social sciences fields who had at least six months of experience in academic research and were currently engaged in research on a specific topic. We recruited 18 participants (12 female, 6 male) with an average age of 27.0 ($\sigma = 4.07$), ranging from 22 to 40. The selected participants ranged from novice researchers who had yet to publish any papers to those with more extensive research experience, having published up to twelve papers. The participants were also engaged in ongoing research, spanning from the initial ideation stage to the final paper-writing stage. A total of 179 post-assessment records were collected. To extract patterns of interest in the citation network from these records, we developed a codebook based on the post-assessment records and interview transcripts, identifying six codes. Two authors independently applied closed coding and assessed inter-rater reliability using Krippendorff's alpha~\cite{de2012calculating}. We refined the code definitions and repeated the process until achieving an alpha of 0.80. This method yielded a Cohen’s Kappa for each code with an average of 0.82 ($\sigma = 0.09$), ranging from 0.70 to 0.94. Any coding discrepancies were resolved through discussions between the authors.
\section{Result}

\subsection{Patterns of Browsing Citation Networks}

\Cref{fig:teaser} illustrates the six patterns identified from our analysis, where each pattern refers to the structures that participants felt interested in while examining citation networks through our system.

We found that individuals adopted either a \textbf{layout-oriented} or \textbf{connection-oriented} perspective when browsing and finding papers of interest. 
First, individuals with a layout-oriented perspective aimed to understand the network's overall \textit{``shape''} rather than each citation link. This involved finding structures such as clusters, layers, and circles through the nodes' positions and the edges' orientations.
In contrast, individuals with a connection-oriented perspective recognized and analyzed each citation connection between nodes, which can be seen as a bottom-up approach. 

We also identified three key elements of interest---Fields, Bridges, and Foundations---for each perspective, thus leading to six distinct patterns of interest and reasons for engagement. 

\textbf{Fields are groups of papers related to a single research topic.} In the layout-oriented perspective, fields corresponded to clusters, where nodes positioned closer and forming a cluster-like structure are assumed to cover similar topics. Participants used clusters as basic units for network exploration and navigated across multiple cluster areas. In the connection-oriented perspective, a field corresponds to a set of nodes with many connections to each other (i.e., a near-complete graph). As opposed to the layout-oriented perspective, where fields are treated as a means of exploration, this perspective views fields as the browsing target. 
For instance, according to P16, if he were to find papers that heavily cite each other, he would expect them to represent a closed research field and consider that he had gained a sufficient number of readings in the field.

\textbf{Bridges are connections relating to two or more papers or topics simultaneously.} In the layout-oriented perspective, this is the area between clusters, and participants often assumed that the distance between a cluster and a node indicated similarity. For instance, P7 chose a paper node \textit{``belonging to the left area but slightly to the right,''} expecting it to cover mainly the topic of the left area with some relevance to the right area. In the connection-oriented perspective, bridges are chains of nodes connecting two papers or fields. Bridges facilitated exploratory browsing of networks as participants' focus naturally expanded to bridges once they found fields of interest. Moreover, for participants developing specific research topics, the browsing centered around finding bridges. They searched for papers that blend two or more topics to find seed papers for a multidisciplinary research topic (P13) or to check for similar existing ideas to verify the novelty of their research idea (P15).

\textbf{Foundations are stages in the overall development of the topic (e.g., pioneering, seminal, or recent).} In the layout-oriented perspective, this was mostly stacks of layers in the rank-based layout. Participants saw each layer as representing a stage in the overall development of the research. In the connection-oriented perspective, reasoning about such temporal development was less straightforward, but nodes with many edges to numerous unspecified nodes were identified as seminal papers.

\subsection{Layout vs. Connection-Oriented Perspectives}

\begin{figure}[t]
 \centering
 \includegraphics[width=\linewidth]{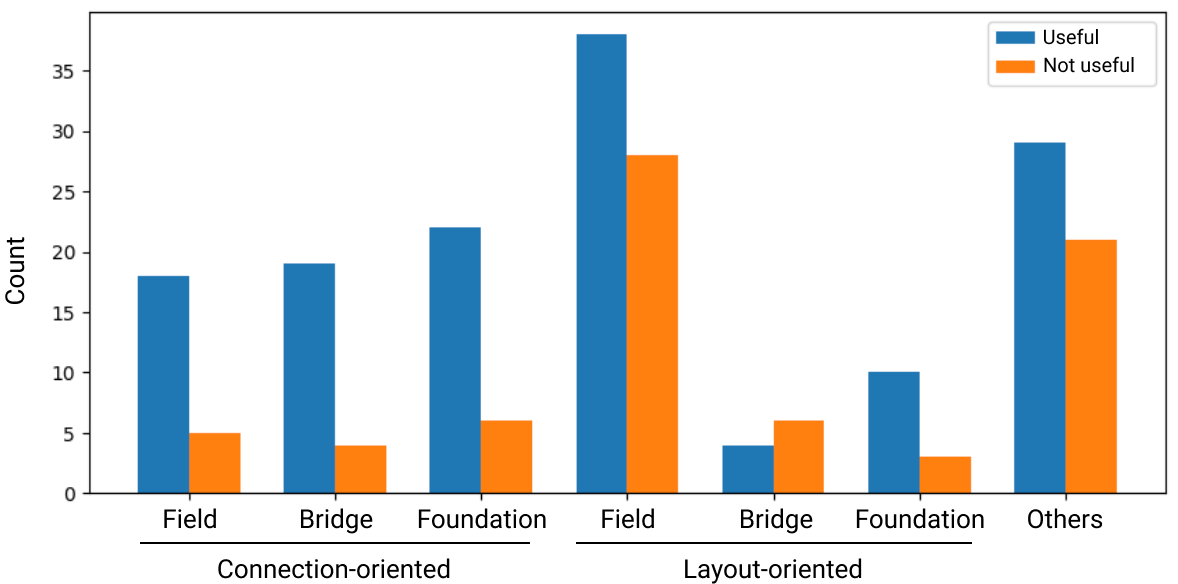}
 \caption{Users' evaluation of the usefulness of each pattern in post-assessment records. Note that a single post-assessment can belong to more than two types.}
 \label{fig:bar-chart}
\end{figure}

Participants generally preferred the layout-oriented perspective but found that papers identified through the connection-oriented perspective were typically more useful. Participants recorded more post-assessment records in the layout-oriented perspective, especially for the Field component (\Cref{fig:bar-chart}). However, post-assessment results revealed that papers discovered from a layout-oriented perspective focusing on Field and Bridge, as well as those categorized as \textit{``Others''} (not belonging to any specific type), had comparable numbers of useful and non-useful records. In contrast, papers explored from a connection-oriented perspective were considered more useful.
Logistic regression was conducted to examine these findings further. The independent variables in the regression model were the binary indications of belonging to each of the six types, and the dependent variable was the binary decision of usefulness, using participant ID as a group factor, as there were ten repeated measures per participant. The results showed significant effects for Bridge and Foundation in connection-oriented and Foundation in layout-oriented perspectives ($p < .05$), with coefficients of 1.37, 1.10, and 1.33, respectively, indicating a strong effect size and suggesting their positive influence in determining usefulness. However, strategies exploring Field and Bridge in layout-oriented showed not only insignificance but also small (layout-oriented Field: 0.44) or even negative coefficient (layout-oriented Bridge: -0.79), suggesting these strategies might not contribute to useful discovery.

A preference for a layout-oriented perspective may arise from the ease of forming an interpretive framework based on the overall layout rather than analyzing numerous connections from the start. Participants often relied on straightforward rules to interpret the layout. Many assumed that closer nodes in a force-directed layout represented more similar topics. This assumption is sometimes correct because papers that heavily cite each other tend to be closer due to the force simulation, but it is not always true. Some participants made even less likely assumptions, such as expecting the innermost part of a cluster to contain older papers and the outermost part to contain newer papers (P16) or interpreting each layer of a rank-based layout as representing similarity to the seed papers (P3, P12). Participants relied on such rules to navigate citation networks and even desired more explicit axes. However, when expectations diverged from the actual data, they felt frustrated. For example, P11 expressed difficulty understanding the layout principles and navigating the network when papers they considered similar and important were scattered rather than clustered in a force-directed layout. 

Among the three layout algorithms, only those with easily inferable construction were preferred. Participants mostly used force-directed layouts and reported they were more intuitive than others, making up about 70\% of post-assessment records in all patterns except layout-oriented Foundation. The circular layout was least preferred, accounting for less than 10\% of records, as the algorithm behind the arrangement of the concentric circles was unclear to most participants.
\section{Discussion}

We investigated what researchers seek to discover in citation networks. Researchers aimed to identify research topics (\textit{Fields}), papers that connect multiple topics (\textit{Bridges}), and the developmental stages within topics (\textit{Foundations}). These components helped them align their research with prior work and define the scope of relevant work. Researchers could approach these components from either a layout-oriented perspective (\textit{i.e.,} interpreting the "shape" of the network) or a connection-oriented perspective (\textit{i.e.,} focusing on actual citations). Those who adopted a layout-oriented approach often relied on intuitive assumptions, such as interpreting proximity in the network as a sign of similarity, and explored the network based on these assumptions without scrutinizing each citation. However, papers identified through layout-based exploration were generally less useful than those found using a connection-oriented approach. This may be due to a potential mismatch between the interpretive frameworks used for network layouts and the inherently connection-based structure of citation networks.

Layouts and connections can interact meaningfully in citation networks. Layouts can offer an overview for exploring various research topics, while citation relationships might provide more accurate guidance, leading to practical outcomes such as identifying new seed papers. Future research could aim to bridge the gap between layout-focused and connection-focused perspectives. One approach could involve designing layouts that align with intuitive assumptions, such as ensuring topic similarity within clusters through semantic embedding~\cite{cohan2020specter}. Adding explicit axes, such as a temporal axis, might also reduce confusion~\cite{shneiderman2006network}. Furthermore, enhancing graph representations by incorporating additional metadata, such as citation counts for nodes and citing contexts~\cite{yousif2019survey} for edges, could improve interpretability. Finally, another approach might focus on detecting and highlighting significant connection relationships within the graphs, enabling users to better understand relationships despite the complexity of the layouts. 

\acknowledgments{This work was partly supported by Institute of Information \& communications Technology Planning \& Evaluation (IITP) grant funded by the Korea government(MSIT) [NO.RS-2021-II211343, Artificial Intelligence Graduate School Program (Seoul National University)] and the National Research Foundation of Korea (NRF) grant funded by the Korean government (MSIT) (No. 2023R1A2C200520911).}

\bibliographystyle{abbrv-doi}

\bibliography{ms}
\end{document}